\documentclass[aps,pre,twocolumn,superscriptaddress,floatfix]{revtex4}
\usepackage{graphicx}
\usepackage[dvips,unicode,colorlinks,linkcolor=blue,citecolor=blue,urlcolor=blue]{hyperref}
\usepackage[english]{babel}
\usepackage{epstopdf}
\begin{document}

\title{One-side surface modifications of graphene}

\author{Alexander V. Savin}
\affiliation{N.N. Semenov Federal Research Center for Chemical Physics,
Russian Academy of Sciences, Moscow 119991, Russia}
\affiliation{Plekhanov Russian University of Economics, Moscow 117997, Russia}

\author{Yuriy A. Kosevich}
\affiliation{N.N. Semenov Federal Research Center for Chemical Physics,
Russian Academy of Sciences, Moscow 119991, Russia}
\affiliation{Plekhanov Russian University of Economics, Moscow 117997, Russia}

\begin{abstract}
We study the dependence of mechanical conformations of graphene sheets located on
flat substrates on the density of unilateral (one-side) attachment of hydrogen,
fluorine or chlorine atoms to them. It is shown that chemically modified
graphene sheet can take four main forms on a flat substrate: the form of a flat sheet located
parallel to the surface of the substrate, the form of convex sheet partially detached from
the substrate with bent edges adjacent to the substrate, and the forms of single and double
rolls on the substrate. On the surface of crystalline graphite, the flat form
of the sheet is lowest in energy for hydrogenation density $p <0.21$, fluorination density
$p <0.20$, and chlorination density $p <0.16$. The surface of crystalline nickel has
higher adsorption energy for graphene monolayer and the flat form of chemically modified sheet
on such substrate is lowest in energy for hydrogenation density $p <0.47$,
fluorination density $p <0.30$ and chlorination density $p <0.21$.
The flat shape of the graphene sheet remains basic on a substrate also when molecular groups CH$_3$,
CH$_2$--CH$_3$ or rings C$_6$H$_5$ are one-side attached to its outer surface. At the attachment density
$ p=1/6 $ (one group per 6 sheet atoms) the sheet becomes the {\it nanocarpet}
the basis of which is formed by a sheet of graphene and the pile of which is formed by the
attached molecular groups forming a tightly packed regular lattice. The addition of hydroxyl
groups OH with attachment density $p=1/4$ leads to the formation of hexagonal lattices of hydroxyl
groups on the outer surface of graphene sheet on a substrate. In this lattice, the groups can form various
configurations of hydrogen bonds, which turns the chemically modified sheet into a multistable system.

\end{abstract}

%\pacs{05.45.-a, 05.45.Yv, 63.20.-e}
\maketitle

%\noindent{\it Keywords}: graphene, substrate, bubble, fold, ripplocation, solitary wave,
%                         molecular dynamics, chain model

\section{Introduction}
\label{Sec1}

Recently, intensive studies were performed of various derivatives of graphene 
(hexagonal monolayer of carbon atoms)
\cite{novoselov04,geim07,soldano10,peng14}, such as graphane CH and fluorographene CF
(a monolayer of graphene, completely saturated on both sides with hydrogen and fluorine)
\cite{sofo07,elias09,nair10}, grafone C$_2$H (graphene monolayer saturated with hydrogen on one
side) \cite{zhou09,balog10,zhao15}, one-side fluorinated graphene C$_4$F \cite{robinson10,enyashin13},
chlorinated graphene C$_4$Cl \cite{sahin12}, and one-side graphene-based polymer carpets \cite{zhang18}.
Valence attachment of an external atom to a graphene sheet leads to local convexity of the sheet
as result of the appearance of the sp$^3$ hybridization at the joining point \cite{ruffieux02,pei2010}.
Therefore, if hydrogen atoms are attached on one side in finite domain of the sheet,
creating a local peace of graphone on the sheet, a characteristic convex deformation
of the sheet occurs in this region \cite{reddy12}. If the hydrogen, fluorine or chlorine atoms are attached
uniformly to the whole one side of the sheet, the whole small sheet will take a convex shape
while the large sheet will fold into a roll \cite{liu13,zhu13,liu15,savin18}.

From the macroscopic point of view, the bending of a graphene sheet under the effect of one-side
chemical modification is similar to the bending of a thin solid film caused by the difference
of surfaces stresses on its sides. Such difference is created during the thin film growth on
a substrate with lattice mismatch \cite{stoney09} or during the one-side epitaxial growth
of a surfactant on the film \cite{martinez90,shell-sor90}. This effect is used, for example,
for the measurements with optical technique of the change in surface stress caused by the
monolayer and sub-monolayer adsorption of a surfactant \cite{shell-sor94,zahl99}.
On the other hand, the equal surface stresses on both sides of elastic thin film cause the change
of the film thickness inversely proportional to its thickness \cite{kos89a,camm89,camm94,kos97},
which can also be used for the characterization of the modification of film surfaces state
induced by surface treatment.

The most convenient way to obtain graphene sheet modified on one side is to attach
the sheet to a flat substrate and further chemically modify its outer surface. For getting
a graphone sheet, the latter needs to be hydrogenated being attached to a substrate.
Modeling of the hydrogenation of a graphene sheet \cite{woellner16,woellner16a} has shown
that the substrate has a significant effect on this process, and is very difficult to obtain
a perfect graphone-like structure with the formula C$_2$H (one hydrogen atom per two carbon atoms).
Hydrogenation leads to the formation of randomly distributed uncorrelated domains with average
hydrogenation density of the sheet $p<0.5$.

Important experimentally observable consequence of the coupling of two-dimensional atomic layer
with elastic substrate is the appearance of the gapped resonance modes in the vibrational spectrum
of the two-dimensional system of distributed oscillators on elastic substrate
\cite{kos89b,gar99,maz15,boech13,beltr14,dahal14}. In Section II we use the value of the spectral
gap in transverse oscillations of the monolayer on elastic substrate for the evaluation
of the coupling strength of carbon, fluorine and chlorine atoms with nickel substrate.

In this paper, we model the dependence of the mechanical conformations (formation of secondary structures)
of graphene sheets, placed on flat substrates (flat surfaces of molecular crystals),
on the density of one-side attachment of hydrogen, fluorine or chlorine atoms. We also consider
mechanical conformations of the graphene sheets with one-side attachment of molecular groups like
CH$_3$, CH$_2$--CH$_3$, C$_6$H$_5$ (benzene ring) or OH (hydroxyl group).

\section{Model of modified graphene sheet}
\label{Sec2}

To model the chemically modified graphene sheet, we use the force field in which
distinct potentials describe the deformation of valence bonds and of valence, torsion and dihedral
angles, and non-valent atomic interactions \cite{savin17}.
In this model, the strain energy of the valence sp$^2$ and sp$^3$ C--C and C--CR bonds,
and of O--H, C--R bonds (here an atom or group of atoms R = H, F, Cl, CH$_3$, CH$_2$--CH$_3$, C$_6$H$_5$,
OH) are described by the Morse potential:
\begin{equation}
V(\rho)=\epsilon_{b}\left[e^{-\alpha(\rho-\rho_0)}-1\right]^2,
\label{f1}
\end{equation}
where $\rho$ and $\rho_0$ are the current and equilibrium bond lengths,
$\epsilon_{b}$ is the binding energy, and the parameter $\alpha$ sets the bond stiffness
$K=2\epsilon_{b}\alpha^2$. The values of potential parameters for various
valence bonds are presented in table \ref{tab1}.
%--------------------- table 1 ------------------------
\begin{table}[bt]
\caption{Values of the Morse potential parameters (\ref{f1}) for different
valence bonds X---Y (C and C$'$ - carbon atoms involved in the formation of the
sp$^2$ and sp$^3$ bonds). \label{tab1}
}
\begin{tabular}{c|cccc}
\hline
 ~~X---Y~~        & ~~$\epsilon_{b}$~(eV)~~ &  ~~$\rho_0$~(\AA)~~ & ~~$\alpha$~(\AA$^{-1}$)~~
 \\
\hline
C---C     &  4.9632 & 1.418 & 1.7889  \\
C---C$'$  &  4.0    & 1.522 & 1.65    \\
C---H     &  4.28   & 1.08  & 1.8     \\
C$'$---F  &  5.38   & 1.36  & 2.0     \\
C$'$---Cl &  3.40   & 1.761 & 2.0     \\
C$'$---O  &  4.01   & 1.41  & 1.65    \\
O---H     &  4.28   & 0.96  & 1.8     \\
\hline
\end{tabular}
\end{table}
%--------------------- table 1 ------------------------
%--------------------- table 2 ------------------------
\begin{table}[tb]
\caption{Values of the parameters of the potential of the valence angle X--Y--Z (\ref{f2}) for different
atoms (atom W = H, F, Cl, O). \label{tab2}
}
\begin{tabular}{c|cccc}
\hline
 ~~X---Y---Z~~& ~~$\epsilon_{a}$~(eV)~~ &  ~~$\varphi_0$~($^\circ$)~~  \\
\hline
C---C---C     &  1.3143 & 120.0 \\
C---C$'$---C  &  1.3    & 109.5\\
C$'$---C$'$---C$'$  &  1.3    & 109.5 \\
C---C---H     &  0.8    & 120.0 \\
C---C$'$---W  &  1.0    & 109.5 \\
C$'$---C$'$---H & 1.0   & 109.5 \\
C$'$---O---H  &  1.0    & 108.5 \\
H---C$'$---H  &  0.7    & 109.5 \\
\hline
\end{tabular}
\end{table}
%--------------------- table 2 ------------------------

Energies of the deformation of the valence angles X--Y--Z are described by the potential
\begin{equation}
U({\bf u}_1,{\bf u}_2,{\bf u}_3)=U(\varphi)=\epsilon_{a}(\cos\varphi-\cos\varphi_0),
\label{f2}
\end{equation}
where the cosine of the valence angle is defined as
$\cos\varphi=-({\bf v}_1, {\bf v}_2)/|{\bf v}_1 || {\bf v}_2 |$,
with vectors $ {\bf v}_1 = {\bf u}_2 - {\bf u}_1 $, ${\bf v}_2 = {\bf u}_3 - {\bf u}_2 $,
the vectors ${\bf u}_1$, ${\bf u}_2 $, ${\bf u}_3 $ specify the coordinates of the atoms forming
the valence angle $\varphi$, $\varphi_0$ is the value of equilibrium valence angle.
Values of potential parameters used for various equilibrium valence angles are presented in table \ref{tab2}.

Deformations of torsion and dihedral angles, in the formation of which edges carbon atoms
with attached external atoms do not participate (torsion angles around sp$^2$ C--C bonds),
are described by the potential:
\begin{equation}
W_1({\bf u}_1,{\bf u}_2,{\bf u}_3,{\bf u}_4)=\epsilon_{t,1}(1-z\cos\phi),
\label{f3}
\end{equation}
where $\cos\phi=({\bf v}_1,{\bf v}_2)/|{\bf v}_1||{\bf v}_2|$, with vectors
${\bf v}_1=({\bf u}_2-{\bf u}_1)\times ({\bf u}_3-{\bf u}_2)$,
${\bf v}_2=({\bf u}_3-{\bf u}_2)\times ({\bf u}_3-{\bf u}_4)$,
the  factor $z=1$ for the dihedral angle (the equilibrium angle $\phi_0=0$)
and  $z=-1$ for the torsion angle (the equilibrium angle $\phi_0=\pi$),
the energy $\epsilon_{t,1}=0.499$ eV (the vectors ${\bf u}_1$,...,${\bf u}_4$ determine
equilibrium positions of the atoms, which form the angle).
More detailed description of the deformation of the torsion and dihedral angles is given in \cite{savin10}.

Deformations of the angles around sp$^3$ bonds C--C$'$, C$'$--C$'$, C$'$--O are described by
the potential:
\begin{equation}
W_2({\bf u}_1,{\bf u}_2,{\bf u}_3,{\bf u}_4)=\epsilon_{t,2}(1+\cos3\phi),
\label{f4}
\end{equation}
with energy $\epsilon_{t,2}=0.03$ eV.

It is worth mentioning that the attachment of two hydrogen atoms on one side of the sheet
to the carbon atoms bonded by valence bond is not energetically favorable  \cite{casolo09}.
Therefore we will consider such attachment configurations on one side of the sheet
of X atoms (X = H, F, Cl, C, O), in which if an X atom is attached to one carbon atom,
the X atoms are no longer attached to the three neighboring carbon atoms.
The valence bonds CX--CX, and the valence and torsion angles formed by
these bonds are absent in such structures.
Therefore the corresponding potentials can be omitted.
%--------------------- table 3 ------------------------
\begin{table}[tb]
\caption{Values of the parameters of the Lennard-Jones potential (\ref{f5})
for different pairs of interacting atoms X, Y.
\label{tab3}
}
\begin{tabular}{c|cccc}
\hline
 ~~X,~~Y~~& ~~$\epsilon_{0}$~(eV)~~ &  ~~$\sigma$~(\AA$^{-1}$)  \\
\hline
C,~~C     &  0.002757 & 3.393 \\
H,~~H     &  0.000681 & 2.471 \\
C,~~H     &  0.001369 & 2.932 \\
F,~~F     &  0.002645 & 3.118 \\
C,~~F     &  0.002700 & 3.256 \\
Cl,~~Cl   &  0.009843 & 3.516 \\
C,~~Cl    &  0.005209 & 3.455 \\
C,~~O     &  0.00658  & 3.025 \\
\hline
\end{tabular}
\end{table}
%--------------------- table 3 ------------------------

The nonvalent van der Waals interactions of atoms are described by the Lennard-Jones potential
\begin{equation}
W_0(r)=4\epsilon_0[(\sigma/r)^{12}-(\sigma/r)^6],
\label{f5}
\end{equation}
where $r$ is the distance between interacting atoms, $\epsilon_0$ is the interaction energy
(equilibrium bond length $r_0=2^{1/6}\sigma$). The used values of the potential parameters
for different pairs of atoms are presented in table \ref{tab3}.
The values of the potential parameters for carbon atoms of the graphene layer are taken
from \cite{setton96}, for the remaining atoms are taken from \cite{rappe92}.

In the simulation, the polarization of the C--F and C--Cl valence bonds was taken into account.
At the atoms forming these bonds, the charges $-q$, $+q$ were used from the PCFF force field
(for the first bond, the charge $q=0.25e$, for the second bond $q=0.184e$).
The interaction of two hydroxyl groups (hydrogen bond OH$\cdots$OH)
was described with the use of the potentials from the PCFF force field.

We define the interaction of the sheet with a flat substrate using the potential $W_s(h)$,
which describes the dependence of the atomic energy on its distance to the substrate plane $h$.
For a flat surface of a molecular crystal, the energy of the interaction of an atom
with a surface  can be described with a good accuracy by the ($k,l$)
Lennard-Jones potential \cite{savin19}:
\begin{equation}
W_s(h)=\epsilon_s[k(h_0/h)^l-l(h_0/h)^k]/(l-k),
\label{f6}
\end{equation}
where $l>k$ is assumed for the exponents. The potential (\ref{f6}) has a minimum
of $W_s(h_0)=-\epsilon_s$ ($\epsilon_s$ is the binding energy of an atom with the substrate).
For a flat surface of crystalline graphite, the exponents in the potential are
$l=10$, $k=3.75$. The binding energy is $\epsilon_s = 0.052$, 0.0187, 0.0465 and 0.1026~eV
for the C, H, F, and Cl atoms, the corresponding equilibrium distances are
$h_0=3.27$, 2.92, 3.24, and 3.435~\AA.

When graphene is located on the surface of crystalline nickel, a stronger
chemical interaction of carbon atoms with the atoms of the substrate occurs.
Therefore, the interaction of the carbon atom in the graphene sheet with the (111) surface
of the Ni crystal is more convenient to describe by the Morse potential:
\begin{equation}
W_s(h)=\epsilon_s\{\exp[-\beta(h-h_0)]-1\}^2-\epsilon_s.
\label{f7}
\end{equation}
For a carbon atom, the interaction energy with the surface is $\epsilon_s = 0.133$~eV \cite{lahiri11}
and the equilibrium distance to the substrate plane is $h_0=2.135$~\AA~\cite{gamo97}.

In result of the interaction of a graphene sheet with a crystal surface, a gap of the magnitude
$\omega_0=240$~cm$^{-1}$ appears at the bottom of the frequency spectrum of transverse
oscillations of the sheet~\cite{dahal14}.
From this we can estimate the harmonic coupling parameter of the interaction of the sheet atom
with the substrate $K_0=\omega_0^2M=41$~N/m ($M$ is a mass of carbon atom),
see also \cite{kos89b}, as well as the value of the parameter
$\beta=\sqrt{K_0/2\epsilon_s}=3.1$~\AA$^{-1}$. For fluorine atom we obtain $\epsilon_s=0.13$~eV,
$h_0=1.655$~\AA, $\beta=3.75$~\AA$^{-1}$, for chlorine atom we obtain $\epsilon_s=0.299$~eV, $h_0=2.115$~\AA,
$\beta=3.17$~\AA$^{-1}$.

In the following we will consider only these two substrate potentials. The first potential describes
weak interaction of the sheet with the substrate while the second potential describes the strong
interaction. Other commonly used substrates (surfaces of crystalline silicon Si,
silicon carbide 6H-SiC, silver Ag and gold Au) are characterized by the intermediate
values of the coupling parameters.

\section{Stationary structures of square sheet}
\label{Sec3}

To find the stationary state of the modified graphene sheet, it is necessary to find the
minimum of the potential energy
\begin{equation}
E\rightarrow\min: \{ {\bf u}_{n}\}_{n=1}^{N},
\label{f8}
\end{equation}
where $N$ is the total number of atoms on the sheet, ${\bf u}_n$ is a three-dimensional vector defining
position of the $n$th atom, $E$ is a total potential energy of the molecular system (given by the sum
of all interaction potentials of atoms of the system).
The minimization problem (\ref{f8}) will be solved numerically by the conjugate gradient method.
Choosing the starting point of the minimization procedure, one can obtain all the main stationary states
of the modified sheet bonded with a flat substrate.

\begin{figure}[tb]
\begin{center}
\includegraphics[angle=0, width=0.905\linewidth]{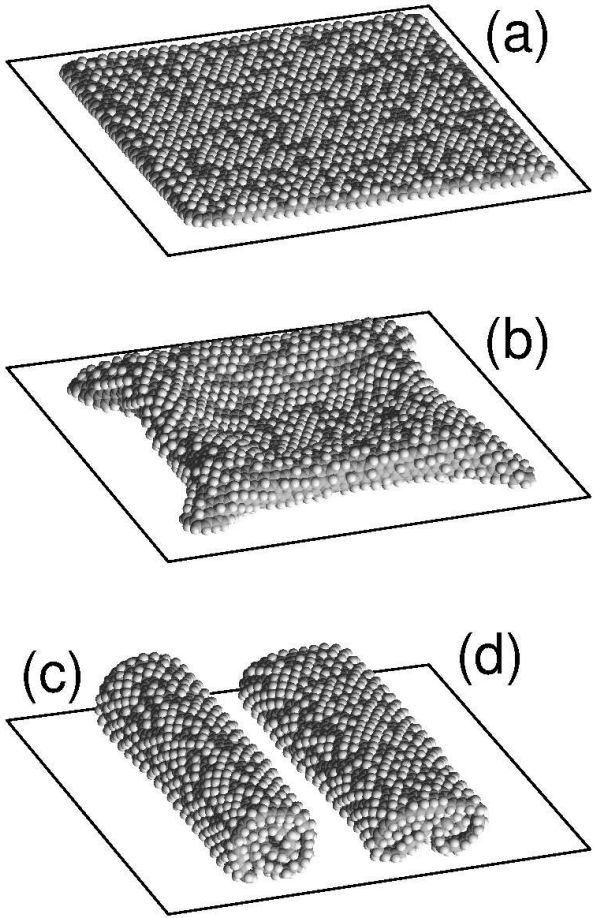}
\end{center}
\caption{\label{fig1}\protect
View of a square graphene sheet of size $8.47\times 8.37$~nm$^2$ (the number of carbon atoms
$N_c=$2798), placed on a flat surface of a graphite crystal with attachment density of hydrogen atoms
to graphene outer surface $p=0.3019$ (the number of hydrogen atoms attached to the surface is $N_h=800$)
with (a) planar structure parallel to the substrate, (b) convex structure with edges attached
to the substrate, and structures of (c) single roll and (d) double roll.
Dark (light) beads show carbon (hydrogen) atoms.
}
\end{figure}

Consider a square graphene sheet of size $8.47\times 8.37$~nm$^2$, consisting of $N_c=2798$
carbon atoms. The sheet has $N_b=148$ edge atoms. To simplify the model, we assume that
only one hydrogen atom is always attached to each edge carbon atom, see fig.~\ref{fig1}~(a).
With maximum one-side hydrogenation (fluorination) of the sheet, $N_m=1324 $ hydrogen atoms can be attached
(for every two internal carbon atoms there is one hydrogen atom).
Thus, the graphene sheet under consideration can be described by the formula
C$_{N_c}$H$_{N_b}$=C$_{2798}$H$_{148}$, and the corresponding graphone sheet by the formula
C$_{N_c}$H$_{N_b+N_m}$=C$_{2798}$H$_{1472}$.
If $ 0\le N_h \le N_m$ of hydrogen atoms are one-side attached to the graphene carbon atoms, the
dimensionless density of the attachment of hydrogen atoms (hydrogenation density) is
$p=N_h/(N_c-N_b)\in [0,0.5]$.

For the modeling of random hydrogenation (fluorination, chlorination) of a graphene sheet,
first we consider the perfect graphone sheet, namely the graphene sheet with $N_m$ hydrogen
(fluorine, chlorine) atoms attached to its outer surface.
Then we randomly remove $N_0$ atoms and get a sheet with dimensionless hydrogenation density
$p=(N_m-N_0)/(N_c-N_b)$. After that, having solved the problem for a minimum of
the energy (\ref{f8}), we find possible stationary structures of the modified sheet.
Each structure (atomic packaging) will be characterized by the specific energy $E_c=E/N_c$.
To evaluate this energy, the removal of $N_0$ atoms will be carried out by 128 independent random ways.
This allows you to find the average value and standard deviation of the specific energy
over 128 independent random implementations of the hydrogenation of the sheet with a fixed
attachment density $p$.
%------------------------------- Fig. 1 ---------------------------------
%----------------------------------------------------------------
%----------------------------- Fig. 2 -----------------------------------
\begin{figure}[tb]
\begin{center}
\includegraphics[angle=0, width=1.0\linewidth]{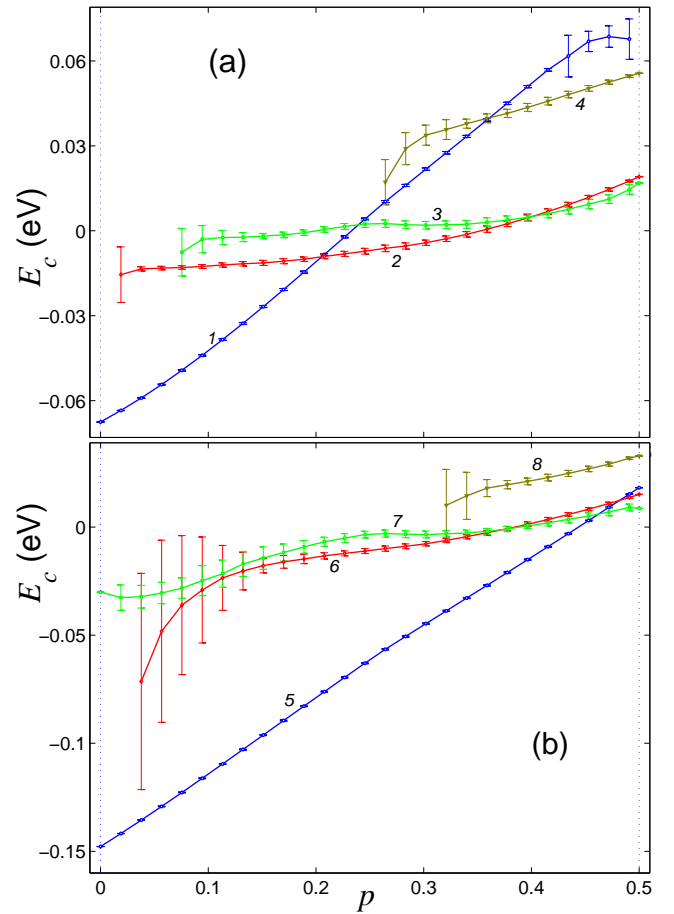}
\end{center}
\caption{\label{fig2}\protect
The dependence of the normalized energy $E_c=E/N_c$ of a square graphene sheet of size
$8.47\times 8.37$~nm$^2$ on the dimensionless hydrogenation density $p$
for a sheet located on a flat surface of (a) graphite crystal and (b)
(111) surface of nickel crystal.
Curves 1, 5 give the dependencies for the flat structure,
curves 2, 6 and 3, 7 -- for the single and double roll structures,
curves 4, 8 -- for partially convex structure with the edges attached to the substrate.
}
\end{figure}
%----------------------------------------------------------------

A free graphone sheet can have two main structures: the single roll and the double roll \cite{savin18}.
Two more stable structures are possible on a flat substrate: a flat form of
a sheet placed parallel to the surface of the substrate and a convex form of the sheet partially
torn off the substrate with folded edges attached to the substrate.
The characteristic appearance of these four stable structures of a sheet on flat substrate
is shown in Fig.~\ref{fig1}. For the flat form, the sheet is located parallel to the substrate
plane and hydrogen atoms are randomly attached to its outer side.
The structure of a single roll has the form of densely-packed roll (scroll) of a sheet with
an external hydrogenated side, lying on a flat substrate.
Double roll structure is realized by folding of the sheet
into two scrolls simultaneously from two opposite edges.

Dependencies of the normalized energy of the sheet $E_c$ on the density of its hydrogenation $p$
for four main structures of a sheet located on the flat surface of a graphite crystal and
on (111) surface of nickel crystal are shown in Fig.~\ref{fig2}. On the surface of graphite,
pure graphene sheet (the hydrogenation density $p=0$) has only one stable flat structure.
Sustainable roll structure can exist only for the hydrogenation density $p>0.018$,
a stable double roll structure can exist only for $p>0.075$, and partially convex structure
can exist only for $p>0.26$.
The flat sheet form remains stable for $p\in [0,0.49]$.
With full hydrogenation (for $p=0.5$) the flat form becomes unstable.
The flat structure is most energetically favorable only for hydrogenation densities $p\in [0,0.21]$,
single roll structure -- for $p\in [0.21,0.41]$, and the double roll structure -- for $p\in [0.41,0.5]$.
Therefore, when a graphene sheet is located on a flat surface of crystalline graphite,
it is hardly possible to achieve its one-sided hydrogenation with density $p>0.21$
because of the folding of the sheet into a roll.

The surface of crystalline nickel has a higher energy of interaction with graphene sheet.
Therefore, the flat form of the graphene sheet remains stable in this case for any density of hydrogenation
of its outer side (for $p\in [0,0.5]$). The flat structure is energetically favorable
for $p<0.47$. Roll folding of the sheet will be stable only for hydrogenation density $p>0.038$,
and for $p\in [0.47,0.5]$ it becomes the most energetically favorable, see Fig.~\ref{fig2}~(b).
Double roll form can exist for any hydrogenation density $p\in [0,0.5]$, but it will always be
energetically disadvantageous. The partially convex sheet structure is always the most
energetically disadvantageous, and it is stable for hydrogenation density $p>0.32$.
Therefore, the location of the graphene sheet on a flat surface
of nickel crystal substrate allows to achieve its hydrogenation with the density close to maximal possible.
%------------------------------- Fig. 3 ---------------------------------
\begin{figure}[tb]
\begin{center}
\includegraphics[angle=0, width=.975\linewidth]{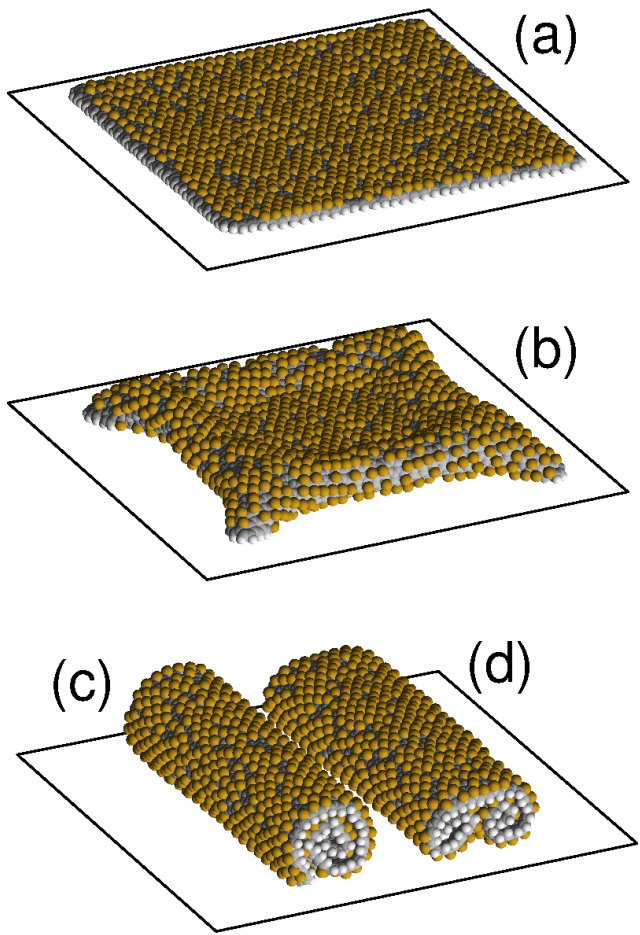}
\end{center}
\caption{\label{fig3}\protect
View of a square graphene sheet
located on flat surface of nickel crystal with density of fluorine atoms attached
to its outer side $p=0.3396$ (the number of fluorine atoms is $N_h=900$)
in (a) planar structure parallel to the substrate, (b) convex structure with the edges attached
to the substrate, and in structures of (c) single roll and (d) double roll.
Gray/white/yellow beads show carbon/hydrogen/fluorine atoms.
}
\end{figure}
%----------------------------------------------------------------
%----------------------------- Fig. 4 -----------------------------------
\begin{figure}[tb]
\begin{center}
\includegraphics[angle=0, width=1.0\linewidth]{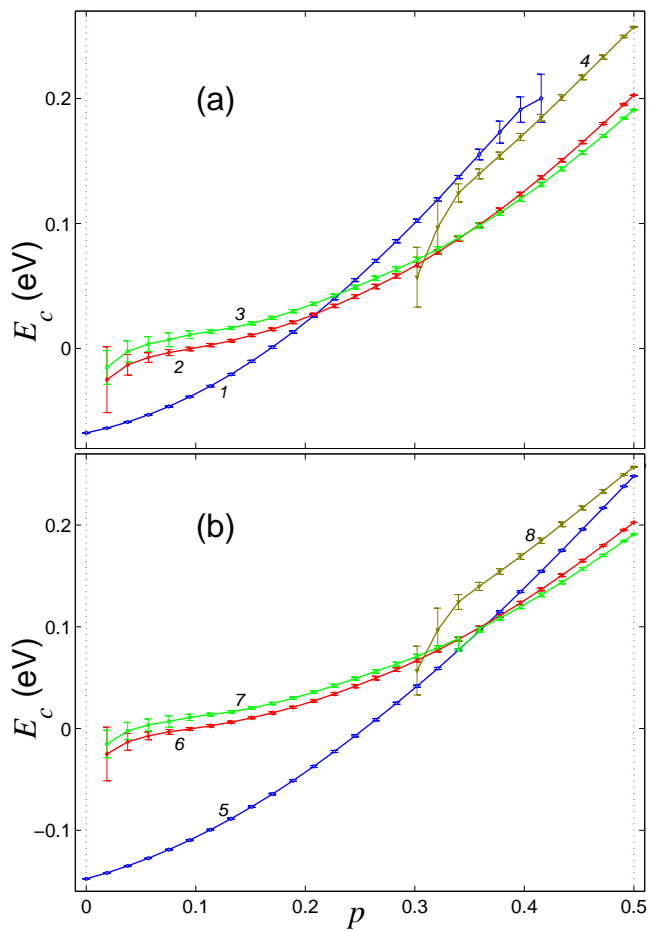}
\end{center}
\caption{\label{fig4}\protect
Dependence of the normalized energy $E_c$ of a square graphene sheet
on the dimensionless fluorination density $p$
for a sheet located on (a) flat surface of graphite crystal and (b)
(111) surface of nickel crystal.
Curves 1, 5 give the dependencies for the flat structure,
curves 2, 6 and 3, 7 -- for the single and double roll structures,
curves 4, 8 -- for partially convex structure with the edges attached to the substrate.
}
\end{figure}
%----------------------------------------------------------------

Similar four stable structures of the graphene sheet are obtained by its unilateral fluorination
(attaching fluorine atoms to the outer surface of the sheet with density $p\in [0,0.5]$).
The characteristic appearance of these four stable structures of fluorinated graphene sheet
on flat substrate is shown in Fig.~\ref{fig3}.
Dependencies of the normalized energy of the sheet $E_c$ on the density of its fluorination $p$
for four main structures of the sheet located on the flat surface of graphite crystal and on (111)
surface of the nickel crystal are shown in Fig.~\ref{fig4}. Stable roll structures can exist
on these substrates only for fluorination density $p>0.018$.

On the surface of crystalline graphite, the planar structure retains its stability for densities
$p\in [0,0.42]$. With higher fluorination density, the flat form becomes unstable.
The partially convex shape of the sheet is stable only for $p\in [0.3,0.5]$.
The flat structure is the most energetically favorable  only for fluorination density
$ p\in [0,0.20]$, the single roll structure -- for density $0.20<p<0.35$, and the double
roll structure -- for density $0.35<p\le 0.5$, see Fig.~\ref{fig4}~(a).
Therefore, when a graphene sheet is located on a flat surface of crystalline graphite, it is impossible
to achieve its one-side fluorination with density $p>0.20$ (this will be prevented by the folding
of the sheet into a roll).
%------------------------------- Fig. 5 ---------------------------------
\begin{figure}[tb]
\begin{center}
\includegraphics[angle=0, width=1.0\linewidth]{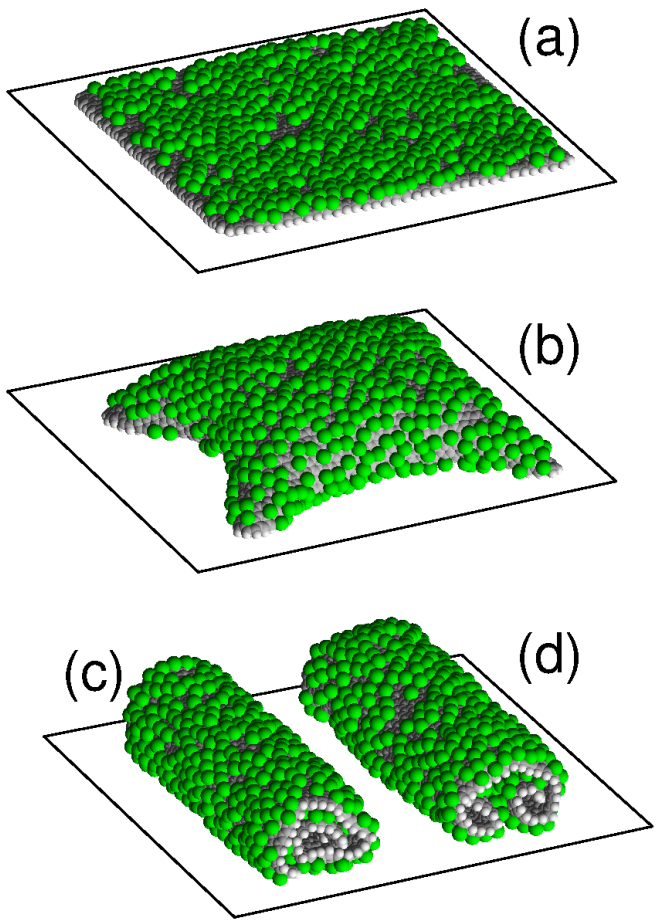}
\end{center}
\caption{\label{fig5}\protect
View of square graphene sheet
located on a flat surface of nickel crystal at attachment density of chlorine atoms
to its outer side $p=0.2075$ (the number of chlorine atoms is $N_h=550$)
with (a) planar structure parallel to the substrate, (b) convex structure with the edges adjacent
to the substrate, and in structures of (c) single roll and (d) double roll.
Gray/white/green beads show carbon/hydrogen/chlorine atoms.
}
\end{figure}
%----------------------------------------------------------------
%----------------------------- Fig. 6 -----------------------------------
\begin{figure}[tb]
\begin{center}
\includegraphics[angle=0, width=1.0\linewidth]{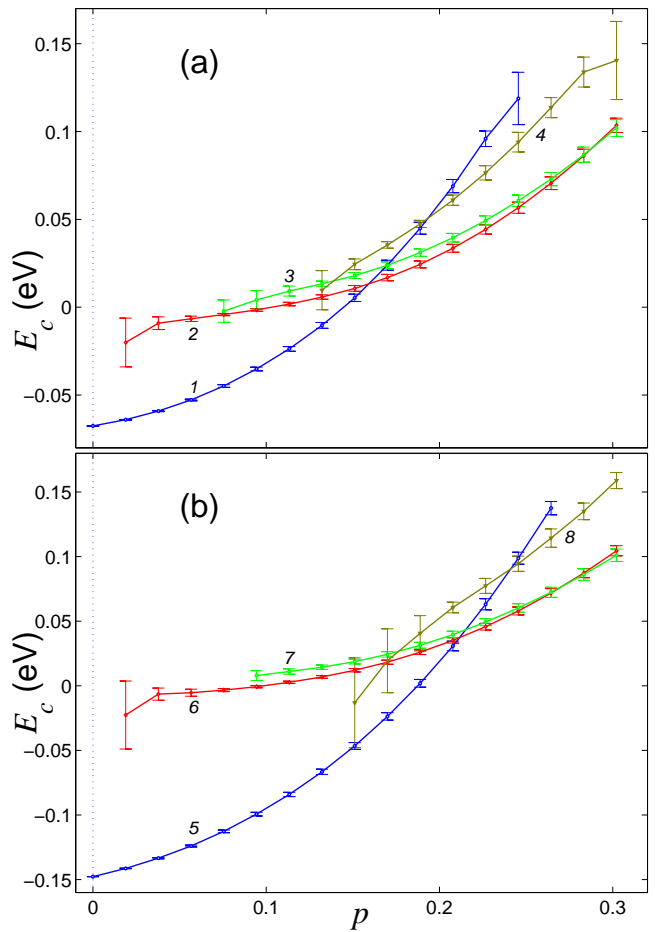}
\end{center}
\caption{\label{fig6}\protect
Dependence of the normalized energy $E_c$ of a square graphene sheet
on the dimensionless chlorination density $p$
for a sheet located on (a) flat surface of graphite crystal and (b)
(111) surface of nickel crystal.
Curves 1, 5 give the dependencies for the flat structure,
curves 2, 6 and 3, 7 -- for the single and double roll structures,
curves 4, 8 -- for partially convex structure with the edges adjacent to the substrate.
}
\end{figure}
%----------------------------------------------------------------

On the surface of crystalline nickel, the flat structure of the sheet remains stable for
$p\in [0,0.5]$, and partially convex form remains stable only for $0.30<p\le 0.5$.
Here, the flat shape is the most energetically favorable for fluorination density $p\in [0,0.30]$,
and the double roll structure -- for $0.30<p\le 0.5$, see Fig.~\ref{fig4}~(b).
Therefore by placing graphene sheet on a flat surface of a nickel crystal, the higher density of
one-side fluorination, $p<0.30$, can be achieved. The decrease of the maximal fluorination density
with respect to the maximal hydrogenation density is related with the larger size of fluoride atom.

Chlorine atoms are larger than the fluorine and hydrogen atoms, what should hinder the
chlorination of graphene.
Characteristic view of four stable structures of the chlorinated graphene sheet on a flat substrate
shown in Fig.~\ref{fig5}.
Dependencies of the normalized sheet energy $E_c$ on the chlorination density $p$ for four main
structures of a sheet located on the flat surface of a graphite crystal and on (111) the surface
of the crystal nickel are shown in Fig.~\ref{fig6}.
On these substrates, a stable single-roll structures can exist only for chlorination density $p\ge 0.019$.

On the surface of crystalline graphite, the planar structure retains its stability for $p\in [0,0.25]$.
With higher chlorination denstities, the flat form becomes unstable.
The double-roll structure is stable only for $p\ge 0.076$, and the partially convex shape
of the sheet is stable for $p>0.13$. The flat structure is the most energy-efficient only
for chlorination densities $p\in [0,0.16]$,
a single-roll structure -- for $0.16<p<0.285$, and the double-roll structure -- for $p>0.285$, see Fig.~\ref{fig6}~(a).
Therefore when a graphene sheet is located on a flat surface of crystalline graphite,
only the one-side chlorination with the density $p<0.16$ can be achieved.

On the surface of crystalline nickel, the flat structure of the sheet remains stable for
$p\in [0,0.264]$. For the higher chlorination, the flat form becomes unstable.
The double-roll structure is stable only for $p\ge 0.094$, and the partially convex shape
of the sheet is stable for $p>0.15$. The flat structure is the most
energy-efficient for chlorination densities $p\in [0,0.21]$, the single-roll structure
-- for $0.21<p<0.264$, and the double-roll structure -- for $p>0.264$, see Fig.~\ref{fig6}~(b).
Therefore for the graphene sheet placed on a flat surface of nickel crystal,
the one-side chlorination can only be achieved for densities $p<0.21$.
%------------------------------- Fig. 7 ---------------------------------
\begin{figure}[tb]
\begin{center}
\includegraphics[angle=0, width=1.0\linewidth]{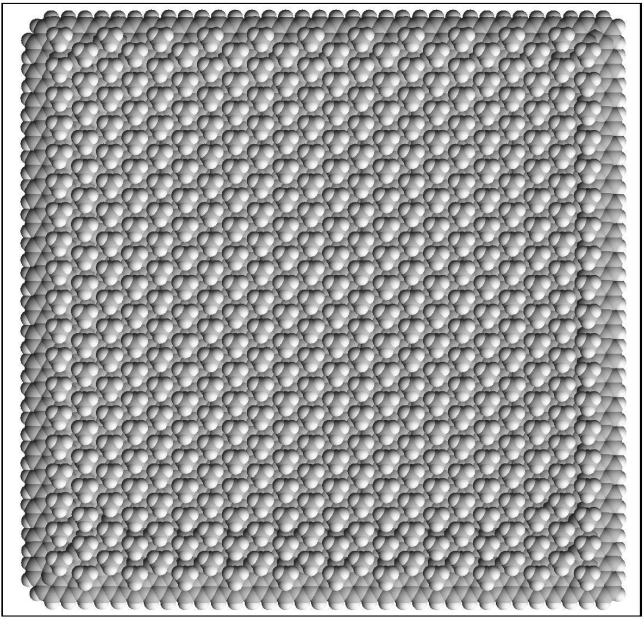}
\end{center}
\caption{\label{fig7}\protect
View of square graphene sheet of size $8.47\times 8.37$~nm$^2$,
placed on a flat surface of graphite crystal
with uniformly attached $N_r=418$ molecular groups R = CH$_3$
(attachment density $p=1/6$).
}
\end{figure}
%----------------------------------------------------------------
%------------------------------- Fig. 8 ---------------------------------
\begin{figure}[tb]
\begin{center}
\includegraphics[angle=0, width=1.0\linewidth]{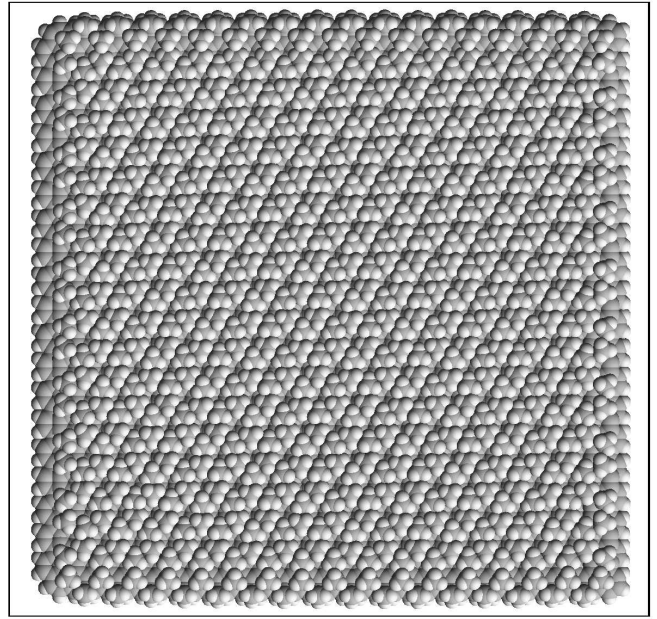}
\end{center}
\caption{\label{fig8}\protect
View of square graphene sheet
placed on a flat surface of nickel crystal with uniformly attached $N_r=660$
molecular groups R = CH$_3$ (attachment density $p=1/4$).
}
\end{figure}
%----------------------------------------------------------------

\section{Graphene-based polymer nanocarpets}
\label{Sec4}

Folding of graphene sheets during their one-side hydrogenation, fluorination or chlorination
is triggered by the transition of some bonds from the sp$^2$ to sp$^3$ form.
If molecular groups $R$ instead of single atoms are attached to one side of the sheet,
the transition of part of the valence bonds to the sp$^3$ form should also lead to the
bending of the sheet. But here the sizes of the attached groups and their interactions
can prevent the folding of the sheet.
At certain attachment density of the groups, a flat polymer {\it nanocarpet} can be formed,
the basis and pile in which are provided, respectively, by the graphene sheet and uniformly attached
polymer molecules $R$.

As an example, we consider graphene sheet with one-side uniformly attached molecular groups,
when the sheet structure can be described by the formula C$_4$R and C$_6$R -- one attached group per
four (attachment density $p=1/4$) and six ($p=1/6$) carbon atoms.
The structures C$_4 $R for R = H, F are considered in detail in \cite{liu15}.
We consider sheets with the attached molecular groups R = CH$_3$, CH$_2 $--CH$_3$, C$_6$H$_5$
(benzene ring) and OH (hydroxyl group).

In the structure of C$_4$R, the attached atoms form a hexagonal
lattice with the distance between adjacent nodes $a=2r_0$, where $r_0=1.418$~\AA~ is the length
of the C--C valence bond in a graphene sheet. In the structure of C$_6$R, the attached atoms
also form the hexagonal lattice, with $a=3r_0$.

Consider a square graphene sheet of size $8.47\times 8.37$~nm$^2$, which consists of
$N_c=2798$ carbon atoms and $N_b=148$ hydrogen atoms attached to the edge carbon atoms.
To model of the structure of C$_6$R, we attach $N_r=418$ molecular groups to the atoms
of the outer side of sheet, and to model the structure of C$_4 $R, we attach $N_r=660$ molecular groups
(see Fig.~\ref{fig7} and \ref{fig8}).
%------------------------------- Fig. 9 ---------------------------------
\begin{figure}[tb]
\begin{center}
\includegraphics[angle=0, width=1.0\linewidth]{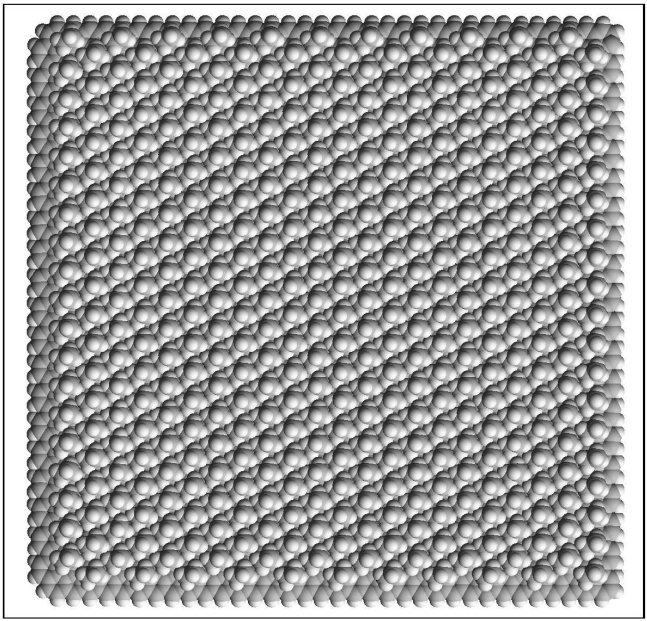}
\end{center}
\caption{\label{fig9}\protect
View of square graphene sheet
lying on a flat surface of graphite crystal with uniformly attached
$N_r=418$ molecular groups R = CH$_2 $--CH$_3$ (attachment density $p=1/6$).
}
\end{figure}
%----------------------------------------------------------------

Upon attachment of atomic groups R = CH$_3$ with density $p=1/6$, the  graphene sheet
lying on a substrate of crystalline graphite retains a flat shape,
and the attached groups form a regular lattice with the distance between neighboring
links $a=4.32$~\AA~, see Fig.~\ref{fig7}.
The small size of this group of atoms allows the existence of a denser structure with $p=1/4$.
For such attachment density, the sheet retains a flat shape only when it is located
on a nickel crystal substrate (and the sheet has the convex shape with folded edges
on the surface of crystalline graphite). Molecular groups on the surface of the sheet form
dense hexagonal structure with lattice period $a=2.97$~\AA~, see Fig.~\ref{fig8}.

\begin{figure}[tb]
\begin{center}
\includegraphics[angle=0, width=1.0\linewidth]{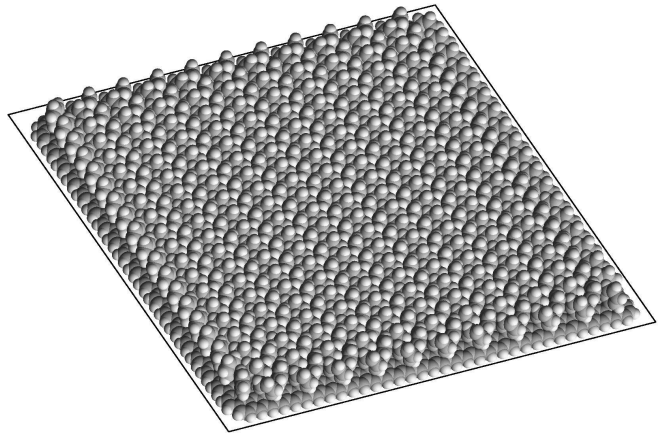}
\end{center}
\caption{\label{fig10}\protect
View of square graphene sheet
lying on a flat surface of graphite crystal with uniformly attached $N_r=418$ molecular groups
R = C$_6$H$_5$ (attachment density $p=1/6$).
}
\end{figure}

Larger sizes of molecular groups R = CH$_2$--CH$_3$, C$_6$H$_5$ allow the attachment to the
sheet only with density $p=1/6$. In this case, the sheet keeps flat
form even being placed on a crystalline graphite substrate, see Fig.~\ref{fig9} and \ref{fig10}.
Here, a lower attachment density of the large-size molecular groups contributes to the conservation of
the flat form of the sheet on the crystalline substrate.

The structure with C$_6$R with R = CH$_2$--CH$_3$ molecular groups (see Fig.~\ref{fig9})
can serve as an example of an ideal polymer nanocarpet, the basis of which is provided
by a graphene sheet and the pile of which is provided by regularly located
double-link fragments of the polyethylene molecule. The distance between neighboring fragments is $a=4.34$~\AA.
A nanocarpet, formed by longer fragments of polyethylene molecules, will have the same structure.
As can be seen from the figure, the attached polymer molecules form
a regular lattice very similar to the monoclinic lattice of crystalline polyethylene.

In the structure of C$_6$R with R = C$_6$H$_5$, the attached benzene rings form the dense crystalline
structure which is formed by the lines of parallel-packed rings, see Fig.~\ref{fig10}.
In neighboring lines, the planes of the rings are inclined one with respect to another by
an angle $\theta\approx 73^\circ$.
%------------------------------- Fig. 10 ---------------------------------
%----------------------------------------------------------------

\section{Graphene sheet with hexagonal lattice of hydrogen bonds}
\label{Sec5}

Consider a graphene sheet of size $8.47\times 8.37$~nm$^2$ lying on a flat surface of
crystalline graphite. To model the structure of C$_4$OH, we attach uniformly 660 hydroxyl groups
to the internal atoms of the sheet. We find stationary states of such structure
(C$_{2798}$H$_{148}$(OH)$_{660}$).
To do this, we find numerically the minimum of the energy (\ref{f8}) using
different initial conditions.

\begin{figure}[tb]
\begin{center}
\includegraphics[angle=0, width=1.0\linewidth]{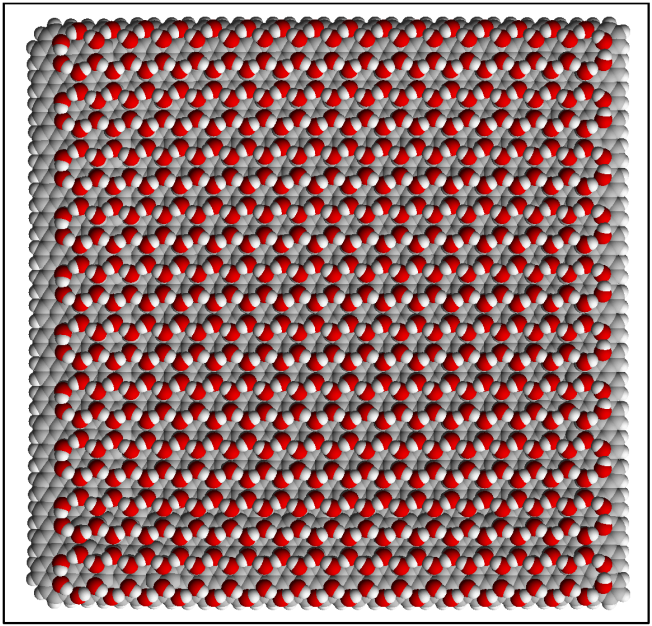}
\end{center}
\caption{\label{fig11}\protect
View of square graphene sheet
lying on a flat surface of graphite crystal
with uniformly attached 660 hydroxyl groups OH (attachment density $p=1/4$).
One of the possible stationary states of the sheet is shown.
}
\end{figure}

A characteristic view of the ground stationary state of the sheet is shown in Fig.~\ref{fig11}.
On a substrate made of crystalline graphite, the sheet always keeps the flat
form parallel to the plane of the substrate. Oxygen atoms on the sheet surface
form a hexagonal lattice with the distance between adjacent nodes $a\approx 3.05$~\AA.
Each non-edge oxygen atom has three nearest-neighbor oxygen atoms.
The hydroxyl group can rotate around the valence sp$^3$ bond C--O, which binds it to the sheet.
The most energetically favorable position of the group is the position in which the O--H bond is
located on the line connecting the oxygen atom with its nearest neighbor, in which case the
hydrogen bond O--H$\cdots$O--H is realized (when the hydrogen bond is formed, the distance between
neighboring oxygen atoms is $a=2.98$~\AA, in the absence of the bond it is $a=3.20$~\AA).
%------------------------------- Fig. 11 ---------------------------------
%----------------------------------------------------------------
%------------------------------- Fig. 12 ---------------------------------
\begin{figure}[t]
\begin{center}
\includegraphics[angle=0, width=1.0\linewidth]{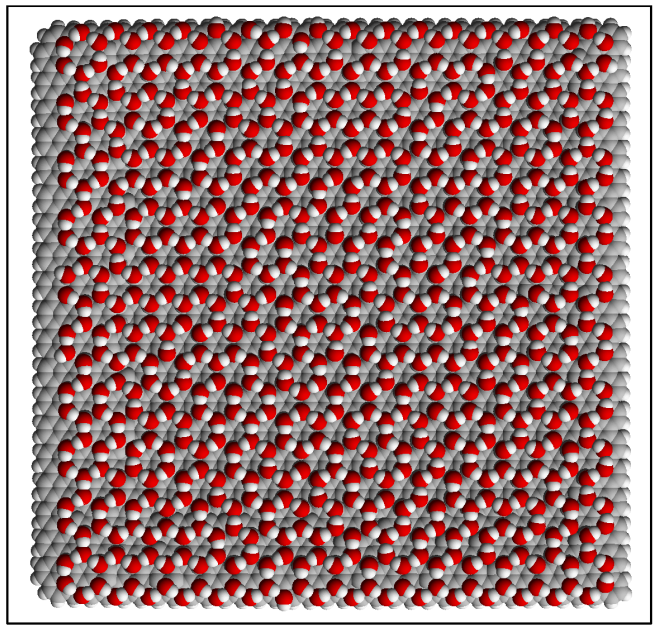}
\end{center}
\caption{\label{fig12}\protect
View of a square graphene sheet
lying on a flat surface of a graphite crystal with uniformly attached  660 hydroxyl groups OH 
(attachment density $p=1/4$). One of the stationary states obtained
with the  modeling of sheet dynamics at temperature $T=300$K is shown.
}
\end{figure}
%----------------------------------------------------------------

In the ground state, the maximal possible number of hydrogen bonds,
equal to the number of hydroxyl groups, should be realized. The following two rules should
be followed in this case: each hydroxyl group should participate in the formation of one hydrogen
bond and each oxygen atom should participate in the formation of two bonds. These two rules can
be implemented in many ways, therefore a graphene sheet with hexagonal lattice of
hydrogen bonds turns to be a multistable system.

To simulate the dynamics of thermalized sheet, the following system of Langevin equations
was numerically integrated:
\begin{equation}
M_n\ddot{\bf u}_n=-\frac{\partial}{\partial{\bf u}_n}E-\gamma M_n\dot{\bf u}_n-\Xi_n,~~~
n=1,...,N
\label{f9}
\end{equation}
where $N=4266$ is the total number of atoms in the molecular structure,
$M_n$ is a mass of the $n$th atom,
${\bf u}_n$ is a three-dimensional vector defining the coordinates of the $n$-th atom,
$\gamma=1/t_r$ is friction coefficient (the relaxation time is $t_r=1$~ps),
$\Xi_n=\{\xi_{n,i}\}_{i=1}^3$ is a three-dimensional vector of normally distributed
random Langevin forces with the following correlation:
$$
\langle\xi_{n,i}(t_1)\xi_{k,j}(t_2)\rangle=2M_nk_BT\gamma\delta_{nk}\delta_{ij}\delta(t_1-t_2)
$$
($k_B$ is Boltzmann constant, $T$ is temperature of the Langevin thermostat).

As an initial condition for the system of equations of motion (\ref{f9}),  we take the
ground stationary state of the sheet shown in Fig.~\ref{fig11}.
Numerical simulation of the dynamics has shown that at temperature $T=300$~K a flat form of
the hydrogenated sheet is always preserved, but the lattice of hydrogen bonds is rearranged,
some bonds are broken, the others are formed. Typical stationary configuration of hydrogen bonds
that is formed on graphene sheet at temperature $T=300$~K is shown in Fig.~\ref{fig12}.
Defects are formed in the lattice of hydrogen bonds.
At the edges of the sheet, some hydroxyl groups do not participate in the formation of bonds,
while inside the sheet some oxygen atoms participate in the formation of only one hydrogen bond,
and some others participate in the formation of three bonds at once.
In comparison with the ground state in in this structure, one additional hydroxyl group
has the surplus energy $\Delta E_{oh}=0.0105$~eV.

\section{Conclusion}
\label{Sec6}

We have performed numerical modeling of mechanical conformations of
graphene sheets placed on flat substrates, caused by
the change in the density of one-side attachment of hydrogen, fluorine or chlorine atoms to the sheet.
It is shown that the chemically modified graphene sheet can take
four main forms on the flat substrate: the form of a flat sheet parallel to the surface of the substrate,
the form of a sheet with convex shape partially detached from the substrate with bent edges attached
to the substrate, and the form of single or double rolls on the substrate.
On the surface of crystalline graphite, the flat sheet form is the most favorable in energy for
hydrogenation density $p<0.21$, fluorination density $p<0.20$ and chlorination density
$p<0.16$. The surface of crystalline nickel has higher energy of graphene adsorption,
here the flat form of chemically modified sheet is the lowest in
energy for hydrogenation density $p<0.47$, fluorination density $p<0.30$ and chlorination
density $p<0.21$. These values of attachment densities can serve for the estimates
of the maximal possible one-side chemical modifications of the flat graphene sheet
(at higher densities, the rolling of the sheet will
prevent it from further chemical modification).
The flat form of the graphene sheet remains main also when molecular groups
CH$_3$, CH$_2$--CH$_3$ or rings C$_6$H$_5$ are unilaterally attached to the outer
surface of the sheet. For attachment density $p=1/6$ (one group per 6 carbon atoms),
the sheet becomes the {\it nanocarpet}, the  basis of which is formed by the graphene sheet
and the pile of which is formed by the attached molecular groups, which
form a tightly packed regular lattice on the surface of the sheet.
Attachment of hydroxyl groups OH with the density $p=1/4$ (one group per 4 carbon atoms)
leads to the formation  of hexagonal lattice of molecular groups on the outer surface
of the sheet. In this lattice, hydroxyl groups can form different chains of
hydrogen bonds OH$\cdots$OH, which turns the modified graphene sheet into a multistable system.
Modeling the dynamics of the sheet at $T=300$K shows that the flat form of the sheet is
resistant to thermal fluctuations which lead only to the thermal rearrangement
of hydrogen bonds in the chains.

\section*{Acknowledgements}
The work was supported by the Russian Science Foundation
(award No. 16-13-10302). The research was carried out using
supercomputers at the Joint Supercomputer Center of the
Russian Academy of Sciences (JSCC RAS).

\end{document}